\documentclass[prl,aps,twocolumn,amsfonts,showpacs,superscriptaddress,floatfix]{revtex4}

\usepackage{epsfig}
\usepackage{psfrag}
\usepackage{color}
\usepackage{graphicx}
\usepackage{subfigure}

\begin{document}

%\twocolumn[\hsize\textwidth\columnwidth\hsize\csname@twocolumnfalse%
%\endcsname

\title{Exciton Hierarchies in Gapped Carbon Nanotubes}
\author{Robert M. KoniK}
\affiliation{Condensed Matter Physics and Material Science Department, Brookhaven National Laboratory, Upton,
NY 11973}

\begin{abstract}
We present evidence that the strong electron-electron interactions in gapped 
carbon nanotubes lead to finite hierarchies
of excitons within a given nanotube subband.
We study these hierarchies by
employing a field theoretic reduction of the gapped carbon nanotube 
permitting electron-electron interactions to be treated exactly.  We analyze this
reduction by employing a Wilsonian-like numerical renormalization group.
We are so able to determine the gap ratios of the one-photon
excitons as a function of the effective strength of interactions.  We also determine
within the same subband the gaps of the two-photon excitons, the single particle gaps, as well as a subset
of the dark excitons.  
The strong electron-electron interactions in addition lead to
strongly renormalized dispersion relations
where the consequences of spin-charge separation can be readily observed.
\end{abstract}

\maketitle

\newcommand{\del}{\partial}
\newcommand{\ep}{\epsilon}
\newcommand{\clsd}{c_{l\sig}^\dagger}
\newcommand{\cls}{c_{l\sig}}
\newcommand{\cesd}{c_{e\sig}^\dagger}
\newcommand{\ces}{c_{e\sig}}
\newcommand{\up}{\uparrow}
\newcommand{\down}{\downarrow}
\newcommand{\il}{\int^{\tilde{Q}}_Q d\la~}
\newcommand{\ilp}{\int^{\tilde{Q}}_Q d\la '}
\newcommand{\ik}{\int^{B}_{-D} dk~}
\newcommand{\ila}{\int d\la~}
\newcommand{\ilpa}{\int d\la '}
\newcommand{\ika}{\int dk~}
\newcommand{\tQ}{\tilde{Q}}
\newcommand{\rh}{\rho_{\rm bulk}}
\newcommand{\ri}{\rho^{\rm imp}}
\newcommand{\sh}{\sig_{\rm bulk}}
\newcommand{\si}{\sig^{\rm imp}}
\newcommand{\rph}{\rho_{p/h}}
\newcommand{\sph}{\sig_{p/h}}
\newcommand{\rp}{\rho_{p}}
\newcommand{\sip}{\sig_{p}}
\newcommand{\drph}{\delta\rho_{p/h}}
\newcommand{\dsph}{\delta\sig_{p/h}}
\newcommand{\drp}{\delta\rho_{p}}
\newcommand{\dsp}{\delta\sig_{p}}
\newcommand{\drh}{\delta\rho_{h}}
\newcommand{\dsh}{\delta\sig_{h}}
\newcommand{\enp}{\ep^+}
\newcommand{\enm}{\ep^-}
\newcommand{\enpm}{\ep^\pm}
\newcommand{\enph}{\ep^+_{\rm bulk}}
\newcommand{\enmh}{\ep^-_{\rm bulk}}
\newcommand{\enpi}{\ep^+_{\rm imp}}
\newcommand{\enmi}{\ep^-_{\rm imp}}
\newcommand{\enh}{\ep_{\rm bulk}}
\newcommand{\eni}{\ep_{\rm imp}}
\newcommand{\sig}{\sigma}
\newcommand{\la}{\lambda}
\newcommand{\ua}{\uparrow}
\newcommand{\da}{\downarrow}
\newcommand{\ed}{\epsilon_d}

Gapped carbon nanotubes are the subject of intense experimental \cite{jim,flu,rmp,wang,torrens,lin} 
and theoretical 
interest \cite{ogawa,axial,levtsv,louie,vasily,kane,vasily1,molinari,louie1,smitha,egger} due 
to their possible application as opto-electronic devices \cite{jim} as well as providing a particularly clean
realization of strongly correlated electron physics in low dimensions \cite{levtsv,egger}.  
Both interests converge
in the study of the tubes' excitonic spectra.  This spectra, due to the effects of a strong Coulomb
interaction in one dimension, is strongly renormalized from what would be expected from the underlying
band structure of the tube.  Its computation therefore requires a non-perturbative approach.

The favoured theoretical approach to studying the excitonic spectra of carbon nanotubes is the use of a Bethe-Salpeter
equation combined with first principle input to approximate the particle and hole wavefunctions
that form the excitons \cite{louie,vasily,vasily1,louie1}.  
This approach has been particularly valuable in that it has allowed a quantitative
description of aspects of excitonic physics, in particular the magnitude of the excitonic gap for the lowest
lying excitons in a given subband of the nanotube.  In this letter we step away from trying to describe
quantitative details of the excitons and instead focus on more qualitative features.  We do
so using an approach that combines a field theoretic reduction of the nanotubes identical to that used
to study Luttinger liquid behaviour in metallic nanotubes with a numerical renormalization group that
enables one to study the effects of gapping out a multi-component Luttinger liquid.  In the process
we find in a regime of strong electron-electron interactions 
a number of new qualitative features to the excitonic spectra (see Fig. 1).

First and foremost we find that a given subband of the nanotube has a finite number of optically active 
one-photon excitons, $\Delta^{1u,i}$,
where the multiplicity depends on the strength of the tube's screened Coulomb interaction.  While
a simple picture of excitons as analogs of the excited states of a hydrogen atom \cite{ogawa}
yields an infinite hierarchy of excitons, we argue this series is truncated to at most
three.
In order to understand
the excitons' binding energy, we also study the single particle spectrum of the tube.  The single particle
gap is strongly renormalized by interactions and can be many multiples of the bare bandgap.  Consequently the
lowest lying
two-excitation continuum, depending on strength of the Coulomb interaction, may be a particle-hole continuum
{\it or} it may be a two-exciton continuum.  This approach treats all excitations of the nanotube on
the same footing.  It thus conflates the difference between an exciton and a bi-exciton.  Indeed interactions
can be sufficiently strong that excitons should then be thought of equally as particle-hole bound states or
bound states of two other excitons.  

\begin{figure}[tbh]
\centerline{\epsfig{file=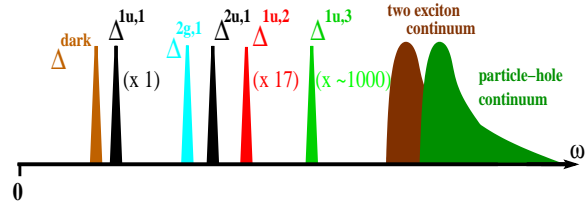,height=1.05in,width=3in}}
\caption{A schematic indicating the position in energy of the various excitons 
together with two-excitation continua
arising from our treatment at larger values of the effective interaction ($K_{c+}^{-1}\geq5$).  
The bracketed numbers
are estimates computed at $K_{c+}=1/7$ of 
the relative absorption strength (relative to the strength of $\Delta^{1u,1}$)
of the three one-photon optically active excitons, $\Delta^{1u,1-3}$.}
\label{cartoon}
\end{figure}

Our results are applicable both to semiconducting nanotubes and 
metallic nanotubes which have been turned semiconducting by the application
of an axial magnetic field \cite{axial}.  In the latter case where the gap is tunable
and can be made much smaller than the bandwidth, we expect our approach to yield results
that are quantitatively accurate.  For tubes that are semiconducting due to band structure,
our results within a one subband approximation
are most reliable for large radius tubes although we believe the qualitative features that we find
to apply equally well to smaller radius tubes.

We focus on a single subband of a carbon nanotube.  To describe this subband at low energies we introduce four
sets (two for the spin, $\sigma$, degeneracy and two for the valley, $\alpha=K,K'$, degeneracy)
of right ($r=+$) and left ($r=-$) moving fermions, $\psi_{r\alpha\sigma}$.  The Hamiltonian
governing these fermions can be written as $H = \int dx (H_{kin} + H_{gap}) + H_{Coulomb}$.
$H_{kin}+H_{gap}$ together give the non-interacting band dispersion, $\epsilon^2(p)=v_0^2p^2+\Delta_0^2$:
\begin{eqnarray}
H_{kin} = -iv_0\psi^\dagger_{r\alpha\sigma}\partial_x\psi_{r\alpha\sigma};~~
H_{gap} = \Delta_0\psi^\dagger_{r\alpha\sigma}\psi_{-r\alpha\sigma},
\end{eqnarray}
where $v_0$ is the bare velocity of the fermions 
and repeated indices are summed.
For the Coulombic part of the Hamiltonian we include only the strongest part of
forward scattering:
$$
H_{Coulomb} = \frac{1}{2}\int dx dx' \rho(x)V_0(x-x')\rho(x'),
$$
where $\rho(x) = \sum_{r\alpha\sigma}\psi^\dagger_{r\alpha\sigma}(x)\psi_{r\alpha\sigma}(x)$.
We also ignore backscattering interactions arising from Coulomb interactions.  
In the presence of a gap term, these
interactions, as they are marginal in the RG sense, are less important.  But as they
lift the model's SU(4) symmetry, they will split
the lowest lying 15-fold multiplet of optically dark excitons.

To study the full Hamiltonian, we treat $H_{gap}$ as a perturbing term (albeit
one to be treated non-perturbatively) of $H_0 \equiv \int dx (H_{kin})+H_{Coulomb}$.  The latter terms
are nothing more than the Hamiltonian of a metallic carbon nanotube.  In excitonic language, we thus
think of $H_{gap}$ as a confining interaction on top of the metallic tube.  The advantage of
doing so is that we are able to treat Coulomb interactions {\it analytically exactly} at the beginning
of the computation.  We do so using bosonization.

If we bosonize $H_0$ in terms of chiral bosons $\phi_{r\alpha\sigma}$
by writing $\psi_{r\alpha\sigma}\sim \exp(i\phi_{r\alpha\sigma})$, 
we arrive at a simple result \cite{kane1,egger}.  The theory
is equivalent to four Luttinger liquids described by the four bosons $\theta_{i}$, $i=c_{\pm},s_{\pm}$
(and their duals $\phi_i$)
\begin{equation}
H_{0} = \int dx\sum_{i}\frac{v_i}{2}\bigg(K_i(\partial_x\phi_{i})^2+K_i^{-1}(\partial_x\theta_i)^2\bigg).
\end{equation}
The four bosons diagonalizing $H_0$ are linear combinations of the original four
bosons and represent an effective charge-flavour separation
where $\theta_{c+} = \sum_{r\alpha\sigma}r\tilde\psi_{r\alpha\sigma}$
is the charge boson and the remaining three bosons reflect the spin, valley, and parity symmetries
in the problem.  The charge boson is the only boson to see the effects of the Coulomb interaction.
Both the charge Luttinger parameter $K_{c+}$ and the charge velocity $v_{c+}=v_0/K_{c+}$ are strongly renormalized.
In particular for long range Coulomb interactions, $K_{c+}$ takes the form
\begin{equation}
K_{c+} = \bigg(1+\frac{8e^2}{\pi\kappa\hbar v_0}\big(\log(\frac{L}{2\pi R}) + c_0\big)\bigg)^{-1/2},
\end{equation}
where $\kappa$ is the dielectric constant of the substrate, $L$ is the length of the nanotube,
$R$ is the tube's radius, and $c_0$ is an $O(1)$ numerical constant dependent on details
of the wrapping vector $(n,m)$ \cite{smitha,egger}.
In typical nanotubes $K_{c+}$ can take on values in the range of $\sim .2$.  With 
an electric field applied transversely to the tube, $K_{c+}$ can easily be made as small as $.1$ \cite{smitha}.
The remaining Luttinger parameters, $K_i$, $i=c_-,s_\pm$ retain their non-interacting values, $1$, and 
so their velocities, $v_i=v_0$ go unrenormalized.  Including the full forward scattering 
leads to a small (upwards) renormalization (on the order of 10\%) of these parameters which
we will not consider here \cite{egger}.  These renormalizations, however, will not change any of our results
qualitatively.

Under bosonization, $H_{gap}$ becomes
\begin{equation}
H_{gap} = \frac{4\tilde\Delta_0}{\pi}(\prod_{i}\cos(\frac{\theta_i}{2}) + \prod_i\sin(\frac{\theta_i}{2})),
\end{equation}
where $\tilde\Delta_0=\Delta_0(\Lambda/v_{c+})^{(1-K_{c+})/4}$ and $\Lambda$ is the bandwidth
of the tube.
This gap term is highly relevant.
Rewriting the gap term in this fashion already gives us important generic features of the excitonic
spectrum.  Firstly, as a perturbation of $H_0$, $H_{gap}$ has
the anomalous dimension, $3/4+K_{c+}/4$.  In turn this implies the
full gaps, $\{\Delta_{\alpha} \}_\alpha$, satisfy the scaling relation
\begin{equation}
\Delta_\alpha = \Lambda (\frac{\Delta_0}{\Lambda})^{4/(5-K_{c_+})}f_\alpha (K_{c+})
\end{equation}
where $\{f_\alpha \}_\alpha$ is a set of a priori unknown dimensionless functions of $K_{c+}$.  We will see that
for the excitons, $f_{Exc}$ is a relatively weak function of $K_{c+}$ while for the single
particle excitations, $f_{sp}$ depends strongly on $K_{c+}$.  This scaling relation tells us
immediately how exciton gaps scale between subbands.  If we take a large radius tube
where the gap of the n-th subband gaps scale as $\Delta_0(n,R) \sim n/R$, the expected
gap ratio between the first and second subband,
is $2^{4/(5-K_{c+})} \sim 1.78$ (this is in rough correspondence to that reported in \cite{flu} and
so provides a straightforward explanation of the ratio problem \cite{kane}).
Secondly, $H_{gap}$ takes the form of a perturbation of a generalized sine-Gordon model (involving
four bosons instead of one).  We thus expect that excitons (within a given subband) will come in hierarchies
whose size is determined by the value of $K_{c+}$ very much like the number of bound states in a sine-Gordon
model, a perturbation on a free boson of the form $\cos(\beta \theta)$, is determined by the parameter $\beta$.

Having completely characterized $H_0$, 
we study the confining effects of $H_{gap}$ 
using the truncated conformal spectrum approach (TCSA) \cite{zamo}
combined with a Wilsonian renormalization group modeled after the numerical renormalization group (NRG)
used to study quantum impurity problems \cite{kon1}.  This methodology permits the study of arbitrary
continuum one dimensional theories provided one can write the theory as the sum of a gapless 
theory, in this case $H_0$, plus a perturbation, here $H_{gap}$.  It is flexible enough that
it can play the same role as DMRG does for lattice models (although as it is an NRG, it accesses
matrix elements more easily).  Among the first uses of the original TCSA, unequipped with an NRG, was to study 
a quantum Ising 
model in a magnetic field \cite{zamo}.  The model we study here, in terms of computational 
complexity, is equivalent
to eight coupled Ising models.  Because of this added complexity, this model could not be studied
using the TCSA without the accompanying renormalization group.  Additional details of this approach
as applied to the problem at hand may be found in Section I of Ref. \cite{epaps}.

Our results for the gaps as a function of $K_{c+}$ are presented in Fig. 2. Details of the analysis
of the numerical data have also been relegated to Ref. \cite{epaps}.  We emphasize again
that these excitations are all found within a single subband of the tube.  We first consider the one-photon
optically active excitons (labelled $\Delta^{1u,i}$, $i=1,2,3$ where here $u$ refers to the antisymmetry
under $\pi-$rotations about the centre of a carbon hexagon -- in bosonic language antisymmetry 
under this rotation corresponds
to antisymmetry under $\theta_{c+,s+}\rightarrow -\theta_{c+,s+}$).  We see that as $K_{c+}$ decreases (i.e. the effective
Coulomb interactions become stronger) the number of one photon excitons grows.  At $K_{c+}\geq 1/5$ the
number of such excitons goes from one to two, while at $K_{c+}\geq 1/6$ the number of such excitons
goes from two to three.  But thereafter this number saturates.  

There are two limits on stable
excitons in the system -- the beginnings of both
the particle-hole continuum (marked in dark green in Fig. 2) and a two-exciton continuum (marked in
maroon in Fig. 2).  
Necessarily exciton gaps cannot cross over these boundaries lest a decay
channel opens to the exciton.  The bottom of the particle-hole continuum is twice the
single particle gap, $2\Delta^{sp}$.  As $\Delta^{sp}$ is a strong function
of $K_{c+}$ (going as $\Delta^{sp}\sim K_{c+}^{-1}$), 
at small $K_{c+}$ there is room for additional excitons.  This is very
much like the sine-Gordon model where as $\beta$ goes to zero, the number of bound states
goes to infinity.  However for an exciton to be stable it must also fall below a two-exciton continuum.
This two-exciton continuum, formed from the lowest single-photon exciton, $\Delta^{1u,1}$, and the lowest 
two-photon exciton, $\Delta^{2g,1}$, is relatively insensitive to $K_{c+}$
and serves to provide a bound on the total number of possible excitons at small $K_{c+}$.

\begin{figure}
\epsfig{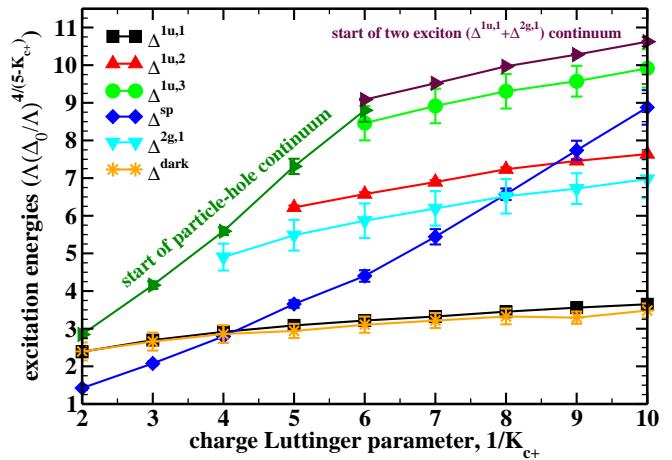}
\caption{The exciton and single particle energies as a function of $K_{c+}$.}
\label{energies}
\end{figure}

In bosonized models considerable intuition can be had by considering the classical limit and asking
what are the classical field solutions that correspond to the excitons/bound states.
This is true here as well.
The first one-photon exciton, $\Delta^{1u,i}$, 
corresponds to a breather-like
solution of the field equations governing $H_0+H_{gap}$.
This solution corresponds to $\theta_{c+}$
interpolating between $0$ to $4\pi$ and back to $0$ while the other three bosons remain fixed at $0$.  It takes the form
\begin{equation}
\theta_{c+}(x,t) = 4\tan^{-1}\frac{\sqrt{1-\omega^2}\cos(\omega t)}{\omega\cosh(\sqrt{1-\omega^2}x)},
\end{equation}
where $\omega$ is related to the (classical) energy of the breather.
We can estimate the energy corresponding to quantizing this classical solution via mean field
theory.
Replacing $H_{gap}$ by $4\tilde\Delta_0/\pi\Xi^3\cos(\theta_{c+}/2)$, where $\Xi\equiv\langle\cos (\theta_f/2)\rangle$
with $\theta_f=\theta_{c-},\theta_{s+}$, or $\theta_{s-}$, we reduce the model
to a pure sine-Gordon theory.  $\Delta^{1u,1}$ is then equal to the gap of this theory's first bound
state \cite{zamo1}
\begin{equation}
\Delta^{1u,1}_{MFT} = 
2\big(2\tilde\Delta_0\Xi^3 \frac{\Gamma(1-\frac{\alpha}{2})}{v_{c+}^{\alpha-1}\Gamma(\frac{\alpha}{2})}\big)^{\frac{1}{2-\alpha}}
\frac{2\Gamma(\frac{\xi}{2})\sin(\frac{\pi\xi}{2})}{\sqrt{\pi}\Gamma(\frac{1}{2}+\xi)},
\end{equation}
where $\alpha=1/4K_{c+}^{2}$ and $\xi=\alpha/(2-\alpha)$.
$\Xi$ is determined self-consistently by replacing $H_{gap}$
with $4\tilde\Delta_0\pi^{-1}\Xi^2\langle\cos(\theta_{c+}/2)\rangle\cos(\theta_f/2)$ and
using \cite{luk} to express $\Xi$ in terms of the effective coupling constant
$4\tilde\Delta_0\pi^{-1}\Xi^2\langle\cos(\theta_{c+}/2)\rangle$.  We consequently find
that as $K_{c+}^{-1}$ varies between $2$ and $10$, $\Delta^{1u,1}_{MFT}/(\Lambda(\Delta_0/\Lambda)^{4/(5-K_{c+})})$ 
varies from $3.1$ to
$4.1$ in reasonable agreement (if shifted upwards)
with the non-perturbative NRG.

The second one-photon exciton, $\Delta^{1u,2}$, has a gap a little more than twice that of $\Delta^{1u,1}$ and
corresponds to
roughly the putative location of $\Delta^{2u,1}$, the first exciton in the second subband 
(which we have argued, by
scaling, is found at $\sim 1.8\Delta^{1u,1}$).  If we were to take into account intersubband Coulomb
interactions, hitherto ignored, $\Delta^{1u,2}$ and $\Delta^{2u,1}$ would mix leading to a set of hybridized
states.  It is thus conceivable that reports of observations of the 
exciton $\Delta^{2u,1}$ are in fact
seeing $\Delta^{1u,2}$ or at least some hybridized combination of the two.

Beyond the one-photon excitons, Fig. 2 also displays the gaps of a number of other excitons.  It shows
the single existing two-photon exciton, $\Delta^{2g,1}$ (here g corresponds to
excitons which are symmetrical
under $\pi$-rotations about a centre of a carbon hexagon).  Unlike the one-photon excitons, 
there is only a single
stable two-photon exciton regardless of the value of $K_{c+}$.  Lying just below $\Delta^{1u,1}$ are found
a set of 15 dark excitons with energy $\Delta^{dark}$.  These excitons 
subsume the dark triplet excitons but are
of greater degeneracy because they not only carry spin quantum numbers but valley quantum numbers as well
(dark excitons with non-trivial valley quantum number were 
termed $K$-momentum excitons in Ref. \cite{torrens}).  As the full symmetry of $H$ is SU(4),
the dark excitons transform as SU(4)'s adjoint.
At twice the energy of $\Delta^{dark}$ begins a two exciton continuum into which any two-photon exciton
will decay.  As $\Delta^{2g,1}$ is just below $2\Delta^{dark}$, we understand this two exciton continuum
essentially precludes any two-photon excitons with energy greater than $\Delta^{2g,1}$.

\begin{figure}
\centerline{\psfig{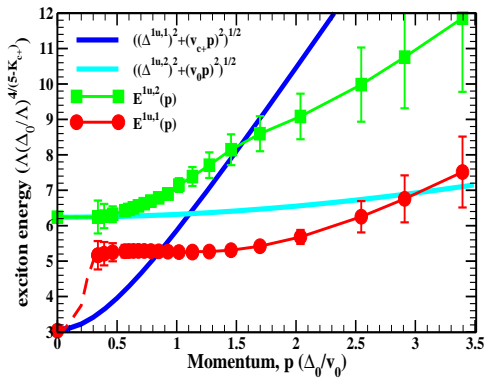}}
\caption{The dispersion of the first two excitons $\Delta^{1u,1}$ and $\Delta^{1u,2}$ for $K_{c+}=1/5$.
Dashed lines in the small, but finite, momentum regions connecting data points are guides to the eye.}
\label{dispersion}
\end{figure}

By examining the momentum dependence of the excitons' dispersions, we can
see the consequences of having two velocities, charge, $v_{c+}$, and flavour $v_0=v_{c-},v_{s\pm}$, 
in the system.  The above semiclassics suggests that the initial dispersion of $\Delta^{1u,1}$, 
$E^{1u,1}(p)$, will behave approximately as $((\Delta^{1u,1})^2+p^2v_{c+}^2)^{1/2}$ (a consequence of
it involving $\theta_{c+}$ degrees of freedom alone).  While we cannot write down the classical
solution corresponding to $\Delta^{1u,2}$, it lies primarily in the spin/flavor sector and so will disperse
governed by the much smaller velocity, $v_0$.
While $\Delta^{1u,2}>\Delta^{1u,1}$, the much
larger charge velocity means that the dispersion of $\Delta^{1u,1}$ will intersect $\Delta^{1u,2}$ at a relatively small
momentum, $p_{int}=(((\Delta^{1u,2})^2-(\Delta^{1u,1})^2)/(v_{c+}^2-v_0^2))^{1/2}$.  
This intersection means the two states will hybridize.
As a consequence of this hybridization, for momenta larger than $p_{int}$, $\Delta^{1u,1}(p>p_{int})$ will disperse
with an effective mass 
determined by $v_0$ whereas $\Delta^{1u,2}(p>p_{int})$ will now vary according to a velocity much closer
to the larger $v_{c+}$.

We observe this general behaviour in our numerical analysis.  Plotted in Fig. 3 are the dispersions
of these two excitons for $K_{c+}^{-1}=5$.  We see that $E^{1u,1}(p)$ increases extremely rapidly 
for momenta up to  
$p\sim \Delta/2v_0$ but thereafter levels out, even decreasing.  This increase is
greater than would be predicted from the relativistic dispersion,
$((\Delta^{1u,1})^2+p^2v_{c+}^2)^{1/2}$, suggesting that charge and spin, even at the smallest momenta,
are intricately linked, leading to an even stronger renormalization of the effective mass. 
If however we combine the renormalization due to $v_{c+}$ with the logarithmic correction
$p^2log(p)$, to the exciton self-energy
predicted in Refs. \cite{kane,vasily1}, we can understand the observed behaviour.
Complementarily, $E^{1u,2}(p)$ initially displays a weak momentum
dependence but then increases with a velocity approximately $v_{c+}/2$ for $p>p_{int}$. 

Finally we touch upon the excitonic signal in absorption spectra.  The strength
of this signal is proportional to the imaginary part of the current-current
correlator.  It is straightforward
to compute the necessary matrix elements of the current operator within the framework
of TCSA+NRG.  We find, as indicated in Fig. 1, that $\Delta^{1u,1}$ has by far the
strongest signal.  While $\Delta^{1u,2}$ should be visible in an absorption
spectra, $\Delta^{1u,3}$, with a weighting 1/1000 of $\Delta^{1u,1}$ is effectively dark
although could be detected indirectly through its effects on the temperature dependence
of the excitonic radiative lifetime \cite{vasily1}.

In conclusion, we have argued for, using a fully many-body, non-perturbative approach, new features
in the excitonic spectrum of gapped carbon nanotubes.  We have
shown that with a given subband, optically active excitons appear in {\it finite} hierarchies.  We have also
shown that the single particle sector experiences extremely strong renormalization so swapping
the beginning of the particle-hole continuum with the two-exciton continuum.  Further work
will include studying the effects of backscattering interactions in the tubes (not treated here)
together with intersubband interactions.
More generally, we have demonstrated a robust method able to treat continuum representations
of arbitrary one-dimensional systems.

RMK acknowledges support from the US DOE
(DE-AC02-98 CH 10886) together with helpful discussions with V. Perebeinos, M. Sfeir, and A. Tsvelik.

%\end{document}
%\documentclass[prl,aps,twocolumn,amsfonts,showpacs,eqsecnum,superscriptaddress,floatfix]{revtex4}

%\documentclass[preprint,doublespace,aps,amsfonts,superscriptaddress,floatfix]{revtex4}

%\usepackage{epsfig}
%\usepackage{psfrag}
%\usepackage{color}
%\usepackage{graphicx}
%\usepackage{subfigure}

%\begin{document}

%\twocolumn[\hsize\textwidth\columnwidth\hsize\csname@twocolumnfalse%
%\endcsname

%\title{Supplementary Material: Exciton Hierarchies in Gapped Carbon Nanotubes}
%\author{Robert M. Konik}
%\affiliation{Condensed Matter Physics and Material Science Department, Brookhaven National Laboratory, Upton,
%NY 11973}

%\maketitle

\section{Supplementary Material: Exciton Hierarchies in Gapped Carbon Nanotubes}

\subsection{Description of NRG treatment of TSA as applied to gapped carbon nanotubes}

We study the confining effects of $H_{gap}$ 
using the truncated conformal spectrum approach.  As orginally developed \cite{zamo},
the TCSA has been used extensively (for a sample consider 
\cite{lassig,klas_mel,takacs,fioravanti,takacs1,takacs2,lepori}).
However here we 
combine it with a Wilsonian renormalization group modeled after the numerical renormalization group (NRG)
used to study quantum impurity problems \cite{kon1}.  
With quantum impurity problems, the NRG exploits the idea that
sites on the Kondo lattice far away from the quantum impurity are less important than sites close to
the impurity in determining the physics.  We have a similar notion here as the perturbation $H_{gap}$ is
relevant in the RG sense (the anomalous dimension of the operator, $3/4+K_{c+}/4$, is less than 2)
and so the eigenstates of the unperturbed Hamiltonian, $H_{0}$,
lowest in energy are most important in determining the physics of the full model.  We are thus
justified in initially truncating the spectrum of $H_0 = \int dx(H_{kin})+H_{Coulomb}$ to a finite number of states (for
the computations chosen here, we initially truncate at $2500$).  To be able to do so, we study the theory at 
finite size, $R_{NRG}$.  In making the system finite, the spectrum becomes discrete with level spacing
$2\pi/R_{NRG}$.  However even at finite size, the conformal nature of $H_0$ means we have complete
control over the spectrum (which can be thought of as a four-fold tensor product 
of states coming from Hilbert spaces of compactified bosons).  

The Hilbert space of the theory can be described more properly as follows.  For a single boson,
with compactification radius, R,
the Hilbert space is built from a set of highest weight states denoted by $|n,m\rangle$
where $n$ gives the center of mass momentum, $n/R$, of the state, and $m$ denotes
the winding number (or charge) of the state \cite{diF}.  $|n,m\rangle$ is related to the true
vacuum state, $|0\rangle$, by
$$
|n,m\rangle = e^{\frac{in}{R}\theta (0) + \frac{imR}{2}\phi(0)}|0\rangle,
$$
where $\phi$ is the boson dual to $\theta$.
Descendant states may be created by acting on the $|n,m\rangle$ with a set of oscillator
modes $a_{-n}$, $\bar{a}_{-n}$, $n>0$, that come from mode expanding the boson.
For our multibosonic system with bosons, $\theta_{c\pm}$, $\theta_{s\pm}$, the set of
highest weight states are formed according to
\begin{eqnarray}
|n_i;m_i\rangle &=& e^{i\sum_j\frac{n_j\sqrt{K_j}}{2}\theta_j (0) + i \sum_j\frac{m_j}{4\sqrt{K_j}}\phi_j(0)}|0\rangle;\cr
\sum_i n_i ~{\rm mod}~ 4 &=& 0; ~~~  \sum_i m_i ~{\rm mod}~ 4 = 0;\cr
&& \hskip -.5in   \sum_i n_im_i ~{\rm mod}~ 4 = 0;\cr
(n_i+n_{i+1}) ~{\rm mod}~ 2 &=& 0; ~~~ (m_i+m_{i+1}) ~{\rm mod}~ 2 = 0.
\end{eqnarray}
These particular conditions upon the highest weight states may be derived by considering the
original bosons that arise in bosonizing the subband fermions (i.e. before the transformation to
the charge and spin bosons).  Descendant states are again formed by acting on these states
with the modes $a_{i,-n}$, $\bar{a}_{i,-n}$, $n>0$, of the bosons.  We note that we can show
this Hilbert space leads to a modular invariant partition function indicating we have a well
defined theory \cite{diF}.  The action of $H_0$ upon a state of the form 
$$
|E^0\rangle = \prod^4_{j=1}(a_{j,-k_{j1}}a_{j,-k_{j2}}\cdots a_{j,-k_{jN_j}})|n_i,m_i\rangle
$$
is 
\begin{eqnarray}
H_0|E^0\rangle &=& \frac{2\pi}{R_{NRG}}\bigg(\sum_j v_j (\frac{K_j n_j^2}{4}+\frac{m_j^2}{16 K_j}-1/12)\cr
&& + \sum_j v_j (\sum^{N_j}_{l=1} k_{jl})\bigg)|E^0\rangle,
\end{eqnarray}
where the $K_j$ are the four Luttinger parameters.

Having described the Hilbert space, we now label the discretized spectrum, 
$|E^0_1\rangle, |E^0_2\rangle,\cdots$ where the spectrum has been ordered in terms of increasing energy.
We then truncate the spectrum keeping the first $N+\Delta$ 
states of the unperturbed theory, $|E^0_1\rangle,\cdots,|E^0_{N+\Delta}\rangle$
(where $N=1500$ and $\Delta=1000$ were typically chosen for this study).  
We further exploit our control over the unperturbed theory
by computing all the matrix elements of $H_{gap}$, i.e. $\langle E^0_i | H_{gap} | E^0_j\rangle$.  This knowledge
then allows us to represent the full Hamiltonian as a $(N+\Delta)\times (N+\Delta)$ matrix which we then diagonalize.  
If $H_0$ were a simple theory, this procedure would be enough to obtain highly accurate
results.  In studying the critical transverse field Ising model perturbed by a parallel magnetic field, Ref. \onlinecite{zamo}
was able to keep only 39 states and yet still obtain the spectrum of the theory to an accuracy
of a few percent.  However a single Luttinger liquid can roughly be thought of (in terms of complexity of the
Hilbert space) as a tensor product of two Ising models.  Thus $H_{kin}+H_{Coulomb}$, being composed of four bosons, 
can be thought
of as equivalent in complexity to a tensor product of eight Ising models.  This eight fold complexity means
that a simple truncation can no longer hope to describe correctly the physics.  This is 
where the numerical
renormalization group (NRG) is crucial, which we now briefly describe.

The NRG begins by taking
the states arising from the diagonalization
of the $(N+\Delta)\times (N+\Delta)$ matrix and ordering them in terms of increasing energy,
$|E^1_1\rangle ,\cdots, |E^1_{N+\Delta}\rangle$.  We then discard the top $\Delta$ states.  
We re-form the truncated Hilbert space by taking the lowest $N$ energy states remaining
from the diagonalization and add to them the next $\Delta$ states from the unperturbed spectrum.
We thus obtain a truncated Hilbert space of the same size ($N+\Delta$): $|E^1_1\rangle,\cdots,|E^1_N\rangle,
|E^0_{N+\Delta+1}\rangle,\cdots,|E^0_{N+2\Delta}\rangle$.  
We re-form the Hamiltonian in this new basis, rediagonalize,
and then repeat the procedure until convergence.  In this way, 
we allow the high energy portion of the unperturbed
spectrum to influence the low energy sector in bite size, numerically manageable ($\Delta$) chunks.

In our study of the exciton gaps, we typically iterate the NRG up to 70 times (and so with a $\Delta=1000$, we
take into account ~$70000$ states of the unperturbed theory).  While the results are not completely converged, we allow
an analytic RG to take over from the numerical RG.  At sufficiently high energies, the flow of a gap
under the RG is governed by a simple one loop equation \cite{kon1}.  We can thus fit the flow equation
to our numerical data at high truncation energies and extrapolate to infinite truncation energy.
For single particle gaps however, we found we needed
to iterate the NRG many more times than for the excitons.  Because
of the large charge velocity (at least at small values of $K_{c+}$), we needed 
to take into account on the
order of 1200000 states from the unperturbed theory.  If we were to use the 
NRG as described in Ref. \cite{kon1},
this would mean iterating the NRG over 1000 times, a too great of numerical burden (the computational
time scales as the number of iterations to the third power).  We have thus adapted the NRG in this
case.  The great majority of the 1200000 states that the NRG needs to run through in the single particle case do
not influence the end result.  We thus run a 'preselection' of the most important states and run the NRG
on those.  This works straightforwardly as follows.

To run the preselection, we run the NRG as initially described for M iterations (and so account
for the first $N+M\Delta$ states of the unperturbed theory).  We have then obtained a reasonable approximate
(though not yet fully converged) of the lowest lying eigenstates of the theory.  Let us for the sake of specificity
focus on the ground state, $|E^M_{GS}\rangle$, obtained after these M-iterations.  Rather than blindly run 
the NRG further, we use
these eigenstates to provide a measure of which of the yet to be iterated through states are important.  Suppose we want
to iterate through a further $K\Delta$ states.  We first compute their overlap with $|E^M_{GS}\rangle$ in second order
perturbation theory, i.e. we compute
$$
\Gamma_i = \frac{|\langle E^M_{GS}|H_{gap}|E^0_{N+M\Delta+i}\rangle|^2}{(E^M_{GS}-E^0_{N+M\Delta+i})},~~~~i=1,\cdots,K\Delta.
$$
We take the $\Gamma_i$'s as a measure of how strongly the state $|E^0_{N+M\Delta+i}\rangle$ will mix into $|E^M_{1}\rangle$ under
the NRG.  Using this measure, we then throw away states, $|E^0_{N+M\Delta+i}\rangle$ with $\Gamma_i$ below some cutoff, $\Lambda$.
We then take the remaining states that have survived this cutoff procedure and run the NRG on those as before.

Typically in applying this procedure to the single particle gaps, we have used $M=70$, $\Delta=1000$, $K$ up to $1100$, and $\Lambda = 2\times 10^{-4}$.
After applying the truncation to the $K\Delta$ states, we were typically 
left with ~70000 states.  Running through these ~70000 states
is then done in several hours rather than the several weeks it would take to run through 
the original $K\Delta$ states.

In analyzing the Hamiltonian we are careful to minimize the necessary computational burden by respecting
the symmetries of the full Hamiltonian.  The full Hamiltonian possesses a $U(1)\times SU(4)$ symmetry.
The $U(1)$ arises from the (valley symmetric) charge degrees of freedom while the $SU(4)$ governs the flavour degrees
of freedom (composed of valley antisymmetric charge degrees of freedom plus spin degrees of freedom).  The
Hamiltonian also possesses $Z_2$ degrees of freedom which arise by transforming 2 of the 4 bosons, $\theta_i$,
via $\theta_i\rightarrow -\theta_i$.  The excitons that are optically active are all $SU(4)$ scalars and
so we run our numerics in this portion of the Hilbert space of the full Hamiltonian.  Similar the single
particle excitations transform as a four dimensional representation (the fundamental under SU(4)) 
while the one branch of dark excitons that
we consider (necessary to understanding the bright exciton spectrum) transforms as a 15 dimensional representation
(the adjoint of SU(4)).  Such large degeneracies of the dark excitons will of course be broken if
we consider the full forward scattering and/or include back scattering terms.  But these interactions are
relatively weak in the dark exciton sector and so the breaking will be comparatively small.   

\subsection{Details of Analysis of NRG Data}

In this section we give some of the details surrounding the interpretation of the NRG data.  In particular we will
give some details on how finite size corrections determine the data analysis.  For $\Delta^{2g,1}$ and $\Delta^{1u,3}$,
finite size corrections are large because both of these excitations are near two-exciton thresholds.

\begin{figure}
\centerline{\psfig{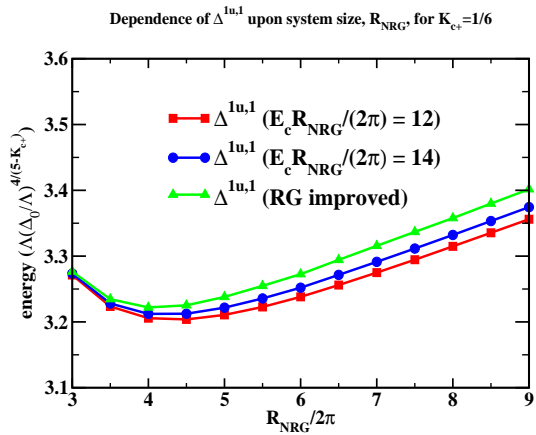}}
\caption{The value of $\Delta^{1u,1}$ for $K_{c+}=1/5$ as function of the system size, $R_{NRG}$, and the cutoff
energy, $E_c$.}
\end{figure}

But we first turn to the easier of the gaps to determine: $\Delta^{1u,1}$, $\Delta^{1u,2}$, and $\Delta^{dark}$.
For specificity we focus on $\Delta^{1u,1}$.  Presented in Fig. 4 is $\Delta^{1u,1}$ as a function
of system size, $R_{NRG}$, for two different truncations of the total number of states (15000 and 75000).
(These truncations are determined by truncating the system so that the energies of the unperturbed
states satisfy $R_{NRG}E/2\pi < 12$ and $R_{NRG}E/2\pi < 14$).  We see that $\Delta^{1u,1}$ possesses only a weak
dependence on $R_{NRG}$.  This, however still leaves the question of which value of $R_{NRG}$ to use for $\Delta^{1u,1}$.
For small values of $R_{NRG}$, we expect the energies of the full theory to reduce to that of the theory's UV fixed point,
i.e. we expect the energies to behave as $1/R_{NRG}$.  We can see this behaviour beginning already at $R_{NRG}=6\pi$.  For
large values of $R_{NRG}$ we expect instead a truncation error (coming from ignoring higher energy states) of the
form $R_{NRG}^{3/2-K_{c+}/2}$.  This error typically leads to an overestimate of the true value of the gap.  The slow increase
of $\Delta^{1u,1}$ that begins at $R_{NRG}=10\pi$ is evidence of this.  Given these behaviors (deviations of $\Delta^{1u,1}$
from its true value), it then makes sense to choose $R_{NRG}$ where $\Delta^{1u,1}$ is at its minimum.  This
we do by choosing $R_{NRG}=9\pi$.  

The one cautionary note we make is that at this value of $R_{NRG}$, $\Delta^{1u,1}$ may still see finite size corrections.
Even though $R_{NRG}$ is sufficiently large so that theory is far enough away from its UV fixed point, $\Delta^{1u,1}$
can see corrections related to virtual processes by which the excitation emits a virtual particle which
then ``travels around the world'' and is reabsorbed.  Such processes are suppressed exponentially in $R_{NRG}$
leading to small corrections to $\Delta^{1u,1}$, i.e. $\delta\Delta^{1u,1} \sim e^{-\alpha R}$.  The coefficient $\alpha$ is either
equal to $\Delta/v_{char}$ where $\Delta$ is one of the other gaps in the theory and $v_{char}$ is its 
characteristic velocity (a so-called
F-term \cite{klas_mel})
or it is related to the binding energy associated with thinking of $\Delta^{1u,1}$ as a bound state of two other excitations (a
so-called $\mu$-term \cite{klas_mel}).
However for $\Delta^{1u,1}$, $\alpha$ (whether arising from an F-term or a $\mu$-term) is sufficiently large so that
the correction should be small (it is difficult to make an absolute statement about this as we cannot easily estimate
the coefficient in front of the exponential which in principle could be large).  
Believing it to be small we make no attempt directly to correct our results for this error.  However dealing
with finite size errors is necessary in studying the gaps $\Delta^{2g,1}$ and $\Delta^{1u,3}$.

We do however do a renormalization group improvement (RGI) of our numerical data 
by taking into account the RG flow that is occurring at fixed $R_{NRG}$.  We see that as we increase
the cutoff energy, $E_c$, the value of $\Delta^{1u,1}$ flows to smaller values.  This flow, at lowest order, is governed by the 
equation
$$
E_c\frac{d\delta \Delta^{1u,1}}{dE_c} = -\beta \delta\Delta^{1u,1},
$$
where $\beta$ here is a function of the anomalous dimension of $H_{gap}$
and $\delta\Delta^{1u,1}=\Delta^{1u,1}(E_c)-\Delta^{1u,1}(E_c=\infty)$.  Integrating this equation, we can express
the value of the gap at finite truncation energy to the gap at infinite truncation energy:
$$
\Delta^{1u,1}(E_c=\infty) = \Delta^{1u,1}(E_c) + \gamma/E_c^{-\beta},
$$
where $\gamma$ is some constant.  We thus can use the data for two different values of $E_c$ to estimate $\gamma$
and so extract the value of $\Delta^{1u,1}$ without cutoff.  The results of doing
so are plotted in Fig. 4 as $\Delta^{1u,1}~({\rm RG~improved})$.

To arrive at the value of $\Delta^{1u,1}$ actually reported in Fig. 2 of the manuscript, 
we assume the RG improved value represents an upper bound
to the value of $\Delta^{1u,1}$ while the value of $\Delta^{1u,1}$ at $E_cR/2\pi = 14$ represents
a lower bound.  It is then the average of these two values that is reported with the error bar
being 1/2 the difference.

\begin{figure}
\centerline{\psfig{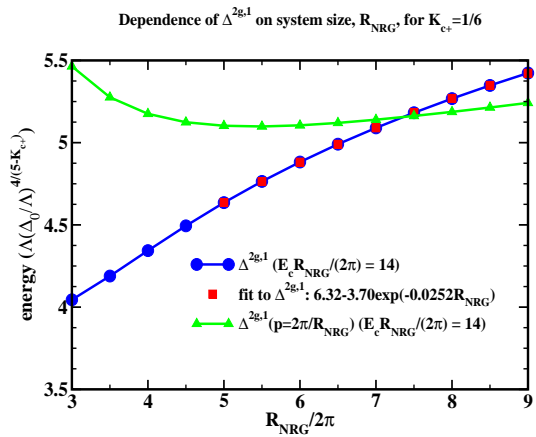}}
\caption{The value of $\Delta^{2g,1}$ for $K_{c+}=1/5$ as function of the system size, $R_{NRG}$, and the cutoff
energy.}
\end{figure}

While $\Delta^{1u,1}$ does not suffer from significant finite size corrections, $\Delta^{2g,1}$ does.  
We show in Fig. 5 $\Delta^{2g,1}$
as a function of $R_{NRG}$.  We see that unlike $\Delta^{1u,1}$, $\Delta^{2g,1}$ steadily increases as a function of $R_{NRG}$.
We understand this to be a consequence of $\Delta^{2g,1}$ being a bound state of either two dark excitons, $\Delta^{dark}$, or
two one-photon excitons, $\Delta^{1u,1}$, and that furthermore 
lies just below these two exciton thresholds, 
i.e $\Delta^{2g,1}$ is only slightly less than $2\Delta^{dark}$ or $2\Delta^{1u,1}$.  In
 such circumstances strong
finite size corrections are expected.  These corrections take the 
form $\delta_{FS}\Delta^{2g,1} \sim \exp(-\alpha R_{NRG})$
where again $\alpha$ is related to the binding energy of the exciton.  
If $\Delta^{2g,1}$ is a bound state of the two dark excitons, $\alpha$ 
is given by
$$
\alpha^2 = (2\Delta^{dark}-\Delta^{2g,1})\frac{1}{2v_0^2},
$$
where $v_0$ appears because the velocity of the dark excitons is very close to $v_0$ as they lie almost entirely
in the spin/flavour sector of the theory.
If instead $\Delta^{2g,1}$ is a bound state of two $\Delta^{1u,1}$, we have instead
$$
\alpha^2 = (2\Delta^{1u,1}-\Delta^{2g,1})\frac{1}{2v^2},
$$
where $v$ here is approximately $v_{c+}$ but will not be exactly that as the exciton $\Delta^{1u,1}$ is influenced
by both the charge and spin sectors of the theory.
Thus if $\Delta^{2g,1}$ is close to either of the two exciton thresholds (or if $v$ is large which
it likely is), $\alpha$ will be small and the finite size corrections, even though exponential, will
decay only slowly as $R_{NRG}$ grows.

To estimate $\Delta^{2g,1}$, we assume the data to have the form 
$\Delta^{2g,1}(R_{NRG})=\Delta^{2g,1}(R=\infty ) + \beta\exp(-\alpha R_{NRG})$.
Fitting the data to this form allows us to estimate $\Delta^{2g,1}(R_{NRG}=\infty)$.  We take this as an upper bound on the value of
$\Delta^{2g,1}(R_{NRG}=\infty)$ as we expect that at larger $R_{NRG}$, $\Delta^{2g,1}$ is overestimated 
in the numerical data because of a finite cutoff, $E_c$.
To find a lower bound, we examine $\Delta^{2g,1}$ as a function of momentum.  Plotted in Fig. 5 
is the numerical
data for $\Delta^{2g,1}(p,R_{NRG})$ where the momentum $p$ is related to $R_{NRG}$ via $2\pi/R_{NRG}$.  
Where $\Delta^{2g,1}(p,R_{NRG})$
and $\Delta^{2g,1}(p=0,R_{NRG})$ meet then serves as a lower bound for $\Delta^{2g,1}$.  
(Incidentally the manner of this
crossing indicates that the dispersion of $\Delta^{2g,1}$ at small momenta is hole-like, 
i.e. $\Delta^{2g,1}(p) \sim \Delta^{2g,1}-p^2/2m_{eff}$.)  
The value for $\Delta^{2g,1}$ that
we plot in Fig. 5 of the manuscript is an average of these upper and lower bounds, while we take
as an uncertainty 1/2 the difference of the two.  

%\begin{references}

%\bibitem{zamo} V. P. Yurov and Al. B. Zamolodchikov, Int. J. Mod. Phys.
%A 6, 4557 (1991).
%\bibitem{lassig}
%M. Lassig, G. Mussardo, and J. L. Cardy, Nucl. Phys. {\bf B 348}, 591 (1991).

%\bibitem{klas_mel}
%T. R. Klassen and E. Melzer, Nucl. Phys. B {\bf 362}, 329 (1991).

%\bibitem{takacs}
%G. Feverati, F. Ravanini, and G. Takacs, Phys. Lett. B {\bf 430}, 264 (1998).

%\bibitem{fioravanti}
%D. Fioravanti, G. Mussardo, and P. Simon, Phys. Rev. E {\bf 63}, 016103 (2000).

%\bibitem{takacs1}
%Z. Bajnok, C. Dunning, L. Palla, G. Takacs, F. Wagner, Nucl. Phys. B {\bf 679}, 521 (2004).

%\bibitem{takacs2}
%G. Takacs, F. Wagner, Nucl. Phys. B {\bf 741}, 353 (2006).

%\bibitem{lepori}
%L. Lepori, G. Mussardo, and G. Toth, J. Stat. Mech. (2008) P09004.

%\bibitem{diF} P. Di Francesco, P. Mathieu, and D. Se\'ne\'chal, {\it Conformal Field Theory},
%Springer-Verlag, New York (1997).

%\bibitem{kon1}
%R. M. Konik and Y. Adamov, Phys. Rev. Lett. {\bf 98}, 147205 (2007).

%\end{references}

\end{document}